\documentclass[aps,pre,reprint,superscriptaddress]{revtex4-1}

\usepackage{amsmath}
\usepackage{color}
\usepackage{graphicx}
\usepackage{txfonts}
\usepackage{enumitem,graphicx,amsmath,amssymb,color,multirow}

\definecolor{dkgreen}{rgb}{0,0.6,0}

\begin{document}

\title{Renormalization of Complex Networks with Partition Functions}

\author{Sungwon Jung}
\affiliation{Department of Physics and Research Institute of Natural Science, Gyeongsang National University, Jinju 52828, Korea}

\author{Sang Hoon Lee}
\email[]{lshlj82@gnu.ac.kr}
\affiliation{Department of Physics and Research Institute of Natural Science, Gyeongsang National University, Jinju 52828, Korea}
\affiliation{Future Convergence Technology Research Institute, Gyeongsang National University, Jinju 52849, Korea}

\author{Jaeyoon Cho}
\email[]{choooir@gmail.com}
\affiliation{Department of Physics and Research Institute of Natural Science, Gyeongsang National University, Jinju 52828, Korea}

\date{\today}

\begin{abstract}

While renormalization groups are fundamental in physics, renormalization of complex networks remains vague in its conceptual definition and methodology.
Here, we propose a novel strategy to renormalize complex networks.
Rather than resorting to handling the bare structure of a network, we overlay it with a readily renormalizable physical model, which reflects real-world scenarios with a broad generality.
From the renormalization of the overlying system, we extract a rigorous and simple renormalization group transformation of arbitrary networks.
In this way, we obtain a transparent, model-dependent physical meaning of the network renormalization, which in our case is a scale transformation preserving the transition dynamics of low-density particles.
We define the strength of a node in accordance with the physical model and trace the change of its distribution under our renormalization process.
This analysis demonstrates that the strength distributions of scale-free networks remain scale-invariant, whereas those of homogeneous random networks do not.

\end{abstract}

\maketitle

\emph{Introduction and summary.}---%
Most core concepts in modern theory of critical phenomena have been built upon renormalization group (RG) transformations~\cite{Fisher1974,Wilson1975,GoldenfeldBook,CandCBook}. 
The key element of RG is the effects of scale transformation~\cite{BarenblattBook} on various system properties. For ordinary systems embedded in Euclidean spaces, {\em rescaling the entire system by a certain factor} is straightforward; 
furthermore, it can also be extended to non-Euclidean spaces, such as the Bethe lattice, which is characterized by its exponentially expanding surface and the uniquely determined pathways between any two sites~\cite{CandCBook}.

In a more general platform of study, commonly referred to as networks~\cite{NewmanBook}, however, the widely accepted framework of RG faces significant challenges. 
A network is essentially a {\em bare-bones} object, the actual manifestations of which are highly diverse. 
This enormous freedom ironically obscures the identification of essential ingredients in RG, especially a (partition) function to preserve in the course of RG transformation. 
As a result, the early excitement on scale-free networks (SFN)~\cite{Holme2019} has turned into uncertainty, failing to suggest a decisive clue to an RG scheme. 
Basically, one has to carefully distinguish between the scale invariance of the degree distribution and that of the system itself~\cite{Stumpf2012}. 
There have been quite a number of attempts to establish and formalize the RG procedure on networks~\cite{BJKim2004,Itzkovitz2005,Song2005,Gfeller2007,Serrano2008,Villegas2023,Nurisso2024,Garuccio2023}. 
Notable examples include the fractal-scaling analysis~\cite{Song2005}, the usage of a hyperbolic-space embedding~\cite{Serrano2008}, focusing on the dynamical property encoded in the graph Laplacian~\cite{Villegas2023,Nurisso2024}, and generating network models achieving the scale-invariant graph partition function~\cite{Garuccio2023}. 
However, a consensus on a standard network RG is yet to be reached.

Arguably, all the earlier schemes circumvent the inherent hurdles by way of {\em redefining} the RG procedure, sometimes incorporating somewhat questionable rules. 
In this paper, we take a different standpoint: 
as a first principle, we keep the well-established rigor of the conventional RG without compromise. 
This is achieved by {\em overlaying} the network with an actual physical model amenable to real-space RG treatment. 
The RG transformation of the underlying network layer is then tied with that of the overlying system. 
Being model-based, this approach makes the motivation and physical meaning of the RG transparent. 
It also means that we abandon the idea of a universal panacea for network RG: 
the utility of a particular RG scheme now relies on the generality of the overlaid system, which we aim to maximize. 
On the technical side, we assert that a faithful network RG transformation should produce a network with real-valued weights. 
In our view, retaining binary weights---merely eliminating or rearranging network nodes and edges---throughout the entire RG process is an oversimplification, losing substantial information.

Our goal is now to seek a physical model fulfilling our requirements. 
As a first step, we consider a {\em quantum} Hamiltonian of free bosons traversing the network nodes, attempting to simulate population flows on a network. 
The partition function of this system and a relevant real-space RG transformation can be rigorously defined. 
As it turns out, the temperature remains invariant under the RG transformation, opening a window to use the scheme in ordinary classical scenarios, with an assumption of high temperature where quantum signatures are washed out. 
Yet, the remaining question is how to extract a weighted network representing this quantum model: 
the problem is that the original parameters in the Hamiltonian lack a natural interpretation applicable in conventional network studies. 
To resolve this, we assume that the bosonic population at each node is observed continuously and strongly, thereby pushing the system further into the classical limit without altering the RG procedure. 
This continuous observation turns our system into a Markov population process~\cite{Kingman1969}. 
Furthermore, with an additional assumption of low population density, we obtain a linear Markov system having a broad application domain. 
At this final stage, we can determine the associated network. 

All in all, our network renormalization can be understood as a scale transformation preserving the transition dynamics of low-density classical particles, where the weight of the edge is interpreted as the transition rate. 
We stress that although we began with a quantum system for technical reasons, the final outcome is fully classical and general. 
For network realizations exhibiting identical behavior to our final system, the RG scheme makes sense. 
In addition, we find that the update rule for the network under the RG transformation is remarkably simple, making our scheme easily accessible.

We test our RG scheme with several illustrative network examples by investigating the probability distribution of the strength, the weighted version of the degree.
We find that static SFNs~\cite{StaticModel,DSLee2004} remain scale-invariant under our RG process even though an extra feature comes in.
Specifically, our RG transformation creates new edges while eliminating some portion of the pre-existing edges.
The pre-existing edges retain the power-law distribution of the strength even though some portion of them are eliminated.
At the same time, newly created edges also display an almost identical power-law distribution.
These two power-law lines coexist in the distribution, remaining almost invariant throughout the RG process after an initial transient phase.
This interesting scale invariance is not observed in the non-scale-free networks we tested.
Thus, our approach would open a new avenue for studying network RG.

\emph{Physical model.}---%
We introduce bosonic mode operators $a_i$ acting on the $i$-th node of the network. 
Every edge is endowed with a hopping rate $J_{ij} = J_{ji}$ (between nodes $i$ and $j$). 
We assume that $J_{ij}$ takes a non-negative real value.
Our Hamiltonian reads
\begin{equation}
    H = - \sum_{i,j=1}^{N} J_{ij} a_i^\dagger a_j - \sum_{i=1}^{N} \mu_i a_i^\dagger a_i \,,
    \label{eq:hamil}
\end{equation}
where $N$ is the number of nodes. 
Here, we introduce site-dependent chemical potential $\mu_i$, which has a negative value to agree with Bose-Einstein statistics. 
The chemical potential plays two roles. 
First, it is a control parameter determining the entire bosonic population together with temperature. 
Second, it enables a real-space renormalization. 
Our RG transformation involves integrating out the degrees at individual nodes. 
Without the local chemical potential, this integration is ill-defined due to the absence of parameters to compensate for local changes.

The partition function is obtained by integrating in the coherent state basis:
\begin{equation}
    \begin{split}
        \mathcal{Z} 
        &= \text{Tr}e^{-\beta H} \\
        &\propto \int \prod_{i=1}^{N} d^2 \alpha_i \langle \alpha_1, \alpha_2, \cdots | e^{-\beta H} |\alpha_1, \alpha_2, \cdots \rangle \\
        &\propto \int \prod_{i=1}^{N} d^2 \alpha_i \exp\left[\beta \left( \sum_{i,j} J_{ij} \alpha_i^* \alpha_j + \sum_j \mu_i |\alpha_i|^2 \right)\right] \,,
    \end{split}
\end{equation}
where $a_i |\alpha_i\rangle = \alpha_i |\alpha_i\rangle$. 
In a single RG step, one of the nodes is eliminated by integrating over the corresponding bosonic mode. 
This procedure leaves the Hamiltonian with updated $J_{ij}$ and $\mu_i$. 
For example, suppose we eliminate node $k$ by integrating over $\alpha_k$. 
As the partition function is a Gaussian integration, one can easily find that it retains the same form, except for $\alpha_k$ absent and the surrounding parameters updated as
\begin{eqnarray}
    \mu_i &\rightarrow& \mu_i - \frac{J_{ik}^2}{\mu_k}\,,\label{eq:Mu}\\
    J_{ij} &\rightarrow& J_{ij} - \frac{J_{ik}J_{jk}}{\mu_k}\label{eq:J}\,.
\end{eqnarray}
This is the simple update rule of our RG transformation. 
The elimination affects all the nodes directly connected to node $k$.
As $\mu_{i}<0$ and $J_{ij}>0$, the RG transformation decreases $|\mu_i|$ and increases $J_{ij}$. 
This is a natural consequence of preserving the whole statistical properties, i.e., the partition function. 
When a node is eliminated, its population moves to the neighboring nodes, meaning that the neighboring chemical potentials decrease in magnitude. 
In addition, the elimination of edges is compensated for by increasing the hopping rates of the neighboring edges.

For network applications, we are tempted to interpret $J_{ij}$ as a transition rate between the associated nodes. 
However, such an interpretation leads to a significant problem. 
In the presence of quantum coherence, transitions occur collectively with enhanced rates. 
Moreover, the transition occurs in the form of an oscillation. 
Thus, $J_{ij}$ has a meaning of a frequency rather than a transition rate. 
We need to wash out this coherence. 
For this, we make a practical assumption that the population of every node is observed continuously and strongly, which happens in the classical world.

To understand the effect of this measurement, consider a simple two-level system described by Hamiltonian
\begin{equation}
    H = - J \sigma_x -\frac{\mu}{2} \sigma_z \,,
\end{equation}
where $\sigma_x$ and $\sigma_z$ are the Pauli operators. 
In the context of our discussion, this system corresponds to the network of two nodes with the total population fixed to one. 
The continuous measurement of the energy level can be described by the following Lindblad equation~\cite{BreuerPetruccioneBook}:
\begin{equation}
    \frac{d\rho}{dt} = -i [H, \rho] + \kappa \left(L \rho L^\dagger - \frac{1}{2} L^\dagger L \rho - \frac{1}{2} \rho L^\dagger L \right) \,,
\end{equation}
where the Lindblad operator is 
\begin{equation}
    L = \frac{1}{2} (I + \sigma_z) \,
\end{equation}
with $I$ being the identity operator. 
Here, the state $\rho$ is conveniently written in the Bloch sphere as
\begin{equation}
    \rho = \frac{I + \vec{r}\cdot\vec{\sigma}}{2} \,,
\end{equation}
where $\vec{\sigma}$ denotes the vector of Pauli operators and the Bloch vector $\vec{r}=(r_x, r_y, r_z)$ with $|\vec{r}| \le 1$ uniquely specifies a two-dimensional density matrix. 
These equations result in the following differential equation:
\begin{equation}
    \frac{d}{dt}
    \left(\begin{matrix}
        r_x \\ r_y \\ r_z
    \end{matrix}\right)
    =
    \left(\begin{matrix}
        -\kappa/2 & \mu/2 & 0 \\ -\mu/2 & -\kappa/2 & J \\ 0 & -J & 0
    \end{matrix}\right)
    \left(\begin{matrix}
        r_x \\ r_y \\ r_z
    \end{matrix}\right) \,.
    \label{eq:diff}
\end{equation}
Provided that the measurement is sufficiently frequent ($\kappa\gg J, |\mu|$), one can handle this equation perturbatively with respect to the small parameters $J/\kappa$ and $\mu/\kappa$. 
In the limit of $J=\mu=0$, Eq.~\eqref{eq:diff} yields the eigenvalues $0$ and $-\kappa/2$. 
The former indicates that the population of each level, associated with $r_z$, is preserved (because $J=0$), while the latter, associated with $r_x$ and $r_y$, indicates that the coherence decays with rate $-\kappa/2$. 
Slightly turning on $J$ and $\mu$, the eigenvalues are altered from $0$ and $-\kappa/2$. 
Here, the actual transition dynamics is governed by the former eigenvalue, which is the smallest in magnitude. 
Again, the other two merely signify the rapid decay of coherence. 
Up to the second-order perturbation, the governing eigenvalue is $-2J^2/\kappa$. 
The interpretation is straightforward. 
The $r_z$ component decays with the rate $2J^2/\kappa$, driving the system into an evenly populated stationary state. 
This two-level transition is identical to a Markov population process with the transition rate $J^2/\kappa$~\cite{Kingman1969}.

Returning to our original discussion, suppose the populations of nodes $i$ and $j$ are $n_i$ and $n_j$, respectively, and a single boson moves from node $i$ to $j$. 
This transition is associated with the off-diagonal element $\langle \cdots, n_i-1, \cdots, n_j+1, \cdots | H | \cdots, n_i, \cdots, n_j, \cdots \rangle$ of Hamiltonian~\eqref{eq:hamil}, which amounts to $J_{ij}\sqrt{n_i (n_j + 1)}$. 
Following the previous analysis, we regard the rate of this transition under continuous measurement as $(n_i + n_i n_j)J_{ij}^2 / \kappa$. 
This is proportional to not only the population at the origin but also the correlation between the populations at the origin and destination. 
This makes sense in scenarios where the population at the destination attracts more visitors. 

Although the emergence of a Markov population process poses potential, the nonlinearity of the transition rate makes its analysis challenging. 
This difficulty is resolved when the population density, i.e., the ratio of the total population to the number of nodes, is very low. 
In this limit, the transition rate approximately becomes $n_i J_{ij}^2 / \kappa$ as the probability of having both $n_i$ and $n_j$ non-zero is negligible. 
This makes the Markov process linear~\cite{Kingman1969}. 
Hereafter, this final system---low-density classical particles subjected to transitions on a network---will be the basis of our network analysis. 
Our RG scheme is applicable to any network system exhibiting the same behavior.

\emph{RG transformation.}---%
We take $w_{ij} \equiv J_{ij}^2$ as the weight of the edge, which constitutes the transition rate of the Markov process.
Note that $J_{ij}^2$ alone does not have the dimension of a rate as the denominator $\kappa$, which determines the time scale, is omitted.
The inverse time scale $\kappa$ is fixed relatively to the initial condition, which we set to one.

We denote by $G_0$ the original unweighted network.
The initial hopping rate $J_{ij}$ in Hamiltonian~\eqref{eq:hamil} is identical to the adjacency matrix of $G_0$.
Without $\mu_i$, the lowest energy is determined by the largest eigenvalue of the matrix $J_{ij}$.
In accordance with Bose-Einstein statistics, the initial $|\mu_i|$ should be larger than this maximum value.
Initially, $\mu_i$ is identical at every node, which we denote by $\mu_0$.

In each RG step, one of the nodes with the largest $|\mu_i|$ is randomly selected and eliminated.
The influence of this elimination spreads out across the neighboring nodes according to the update rule in Eqs.~\eqref{eq:Mu} and \eqref{eq:J}.
Here, the notion of neighboring nodes should be understood with caution.
In weighted networks, there is no clear-cut distinction between connected and disconnected nodes, as an edge with an extremely small weight is physically indistinguishable from the absence of an edge.
The update rule in Eq.~\eqref{eq:J} indicates that the RG process prolifically creates new edges with small weights, quickly saturating the network in the sense that all pairs of nodes have edges with non-zero weights, albeit mostly very small.
This saturation of the network becomes a bottleneck in the course of the numerical calculation, as the exponentially many weights should be stored in memory.
Further RG steps decrease the size of the network, relaxing this demand.
For later analysis, we denote the network under renormalization in the logarithmic unit of the remaining fraction of nodes.
Specifically, we denote by $G_k$ the renormalized network with $N/2^k$ remaining nodes.

\emph{Network analysis.}---%
Our primary interest is in examining how scale-free networks behave under the RG transformation.
We employ the static SFN model~\cite{StaticModel,DSLee2004} as an ideal testbed.
The static SFN was initially introduced to investigate the load or betweenness distribution of SFNs characterized by a power-law degree distribution.
Later, the model has gained popularity as a handy tool to generate a family of SFNs with the asymptotically power-law degree distribution $p(k) \sim k^{-\gamma}$ with the adjustable degree exponent $\gamma \in (2,\infty)$. 
The term ``static'' reflects that the number $N$ of nodes is fixed from the beginning.
Apart from the adjustable degree exponent, the static model has an additional advantage: 
unlike the growing type of SFNs, such as the Barab{\'a}si-Albert (BA) model~\cite{BAmodel}, exhibiting nontrivial degree-degree correlations~\cite{Newman2002} caused by the temporal order of attachment, the static model does not suffer from such an effect.

\begin{figure}
    \includegraphics[width=0.86\columnwidth]{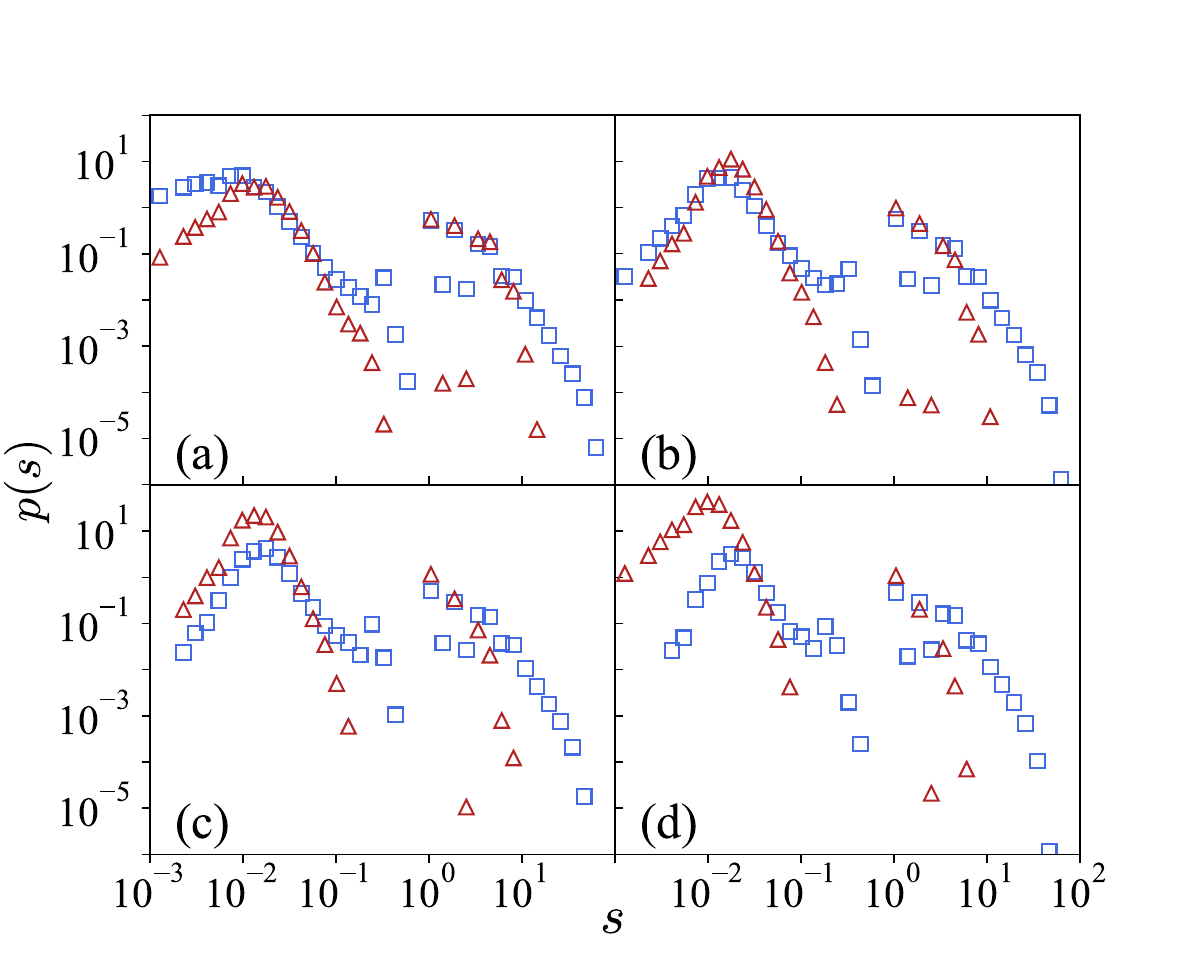}
    \caption{(a)--(d) The distribution $p(s)$ of strength $s$ for the renormalized networks $G_1$--$G_4$ in different scales, respectively, averaged over $10^2$ samples, where the initial network $G_0$ is the static scale-free network with about $10^4$ nodes for $\gamma=3$ (blue squares) and $\gamma=5$ (red triangles). The initial chemical potential is $\mu_0 = -30$.}
    \label{fig:RGed_static_model_strength_distribution}
\end{figure}

\begin{figure}
    \includegraphics[width=0.86\columnwidth]{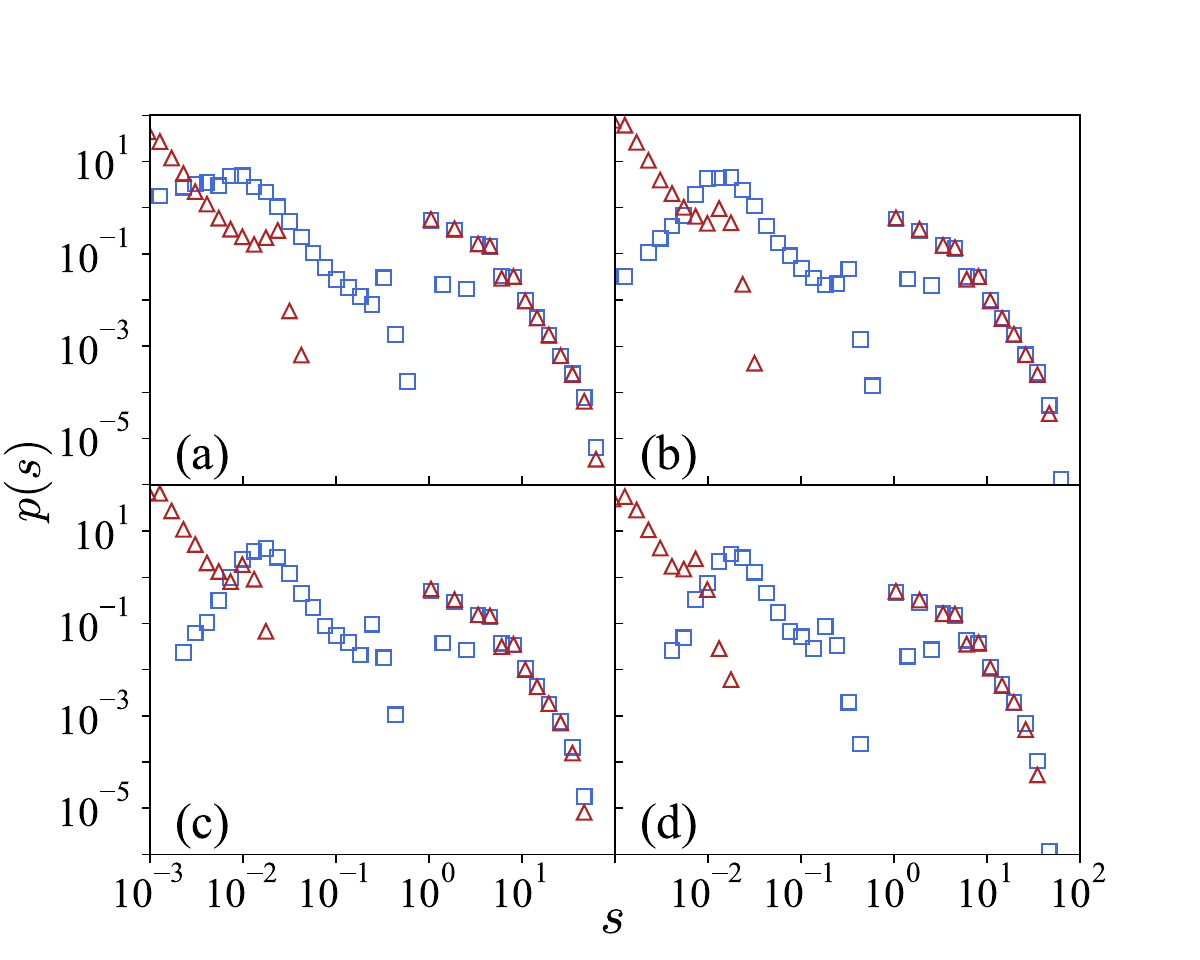}
    \caption{(a)--(d) The same analysis as in Fig.~\ref{fig:RGed_static_model_strength_distribution} with $\gamma=3$, and $\mu_0 = -30$ (blue squares) or $\mu_0 = -120$ (red triangles).}
    \label{fig:different_mu}
\end{figure}

Figs.~\ref{fig:RGed_static_model_strength_distribution} and \ref{fig:different_mu} show our numerical results for the static SFN.
The main quantity of interest is the {\em strength} of nodes~\cite{Barrat2004}, defined as
\begin{equation}
    s_i = \sum_{j} w_{ij} = \sum_{j} J_{ij}^2 \,
    \label{eq:strength}
    \end{equation}
for node $i$.
In the original unweighted network $G_0$, the strength is reduced to the degree $k$.

We generate $10^2$ samples of the network $G_0$ with approximately $10^4$ nodes and obtain the renormalized networks $G_1$ to $G_4$ in different scales during the RG process. We take only the giant component of the generated network samples, so the number of nodes in $G_0$ is slightly reduced: we start from $10\,240$ nodes and $30\,720$ edges, and the average number of nodes left in the giant component is $10\,061.19$ for $\gamma = 3$ and $10\,189.46$ for $\gamma = 5$.
Fig.~\ref{fig:RGed_static_model_strength_distribution} shows the distribution $p(s)$ of strength $s$ for the initial degree exponents $\gamma=3$ and $5$.
They demonstrate the signature of scale invariance in an interesting way.
Following Eq.~\eqref{eq:J}, newly created edges have a weight much smaller than one, while the initially existing edges undergo a minor update of the weight.
Consequently, $s=1$ is the border between the contributions from the new and pre-existing edges.
Fig.~\ref{fig:RGed_static_model_strength_distribution} demonstrates that the distribution of the strength for the newly created edges alone ($s<1$) exhibits the power-law distribution almost identical to the one from the original network.
The power-law distribution in the upper part ($s>1$) results from the contribution by the pre-existing edges {\em excluding} those eliminated during the RG process, which is also remarkable.
The deviation for small $s$ is a finite-size effect.

We also examine the dependency of the result on the chemical potential.
Fig.~\ref{fig:different_mu} shows the similar distributions as in Fig.~\ref{fig:RGed_static_model_strength_distribution} except for the chemical potential chosen as $\mu_0=-30$ or $-120$ while $\gamma=3$ is fixed.
For $\mu_0=-120$, the $s<1$ part is shifted to the left as newly created edges have smaller weights [see Eq.~\eqref{eq:J}].
Otherwise, the scale invariance persists.

\begin{figure}
    \includegraphics[width=0.86\columnwidth]{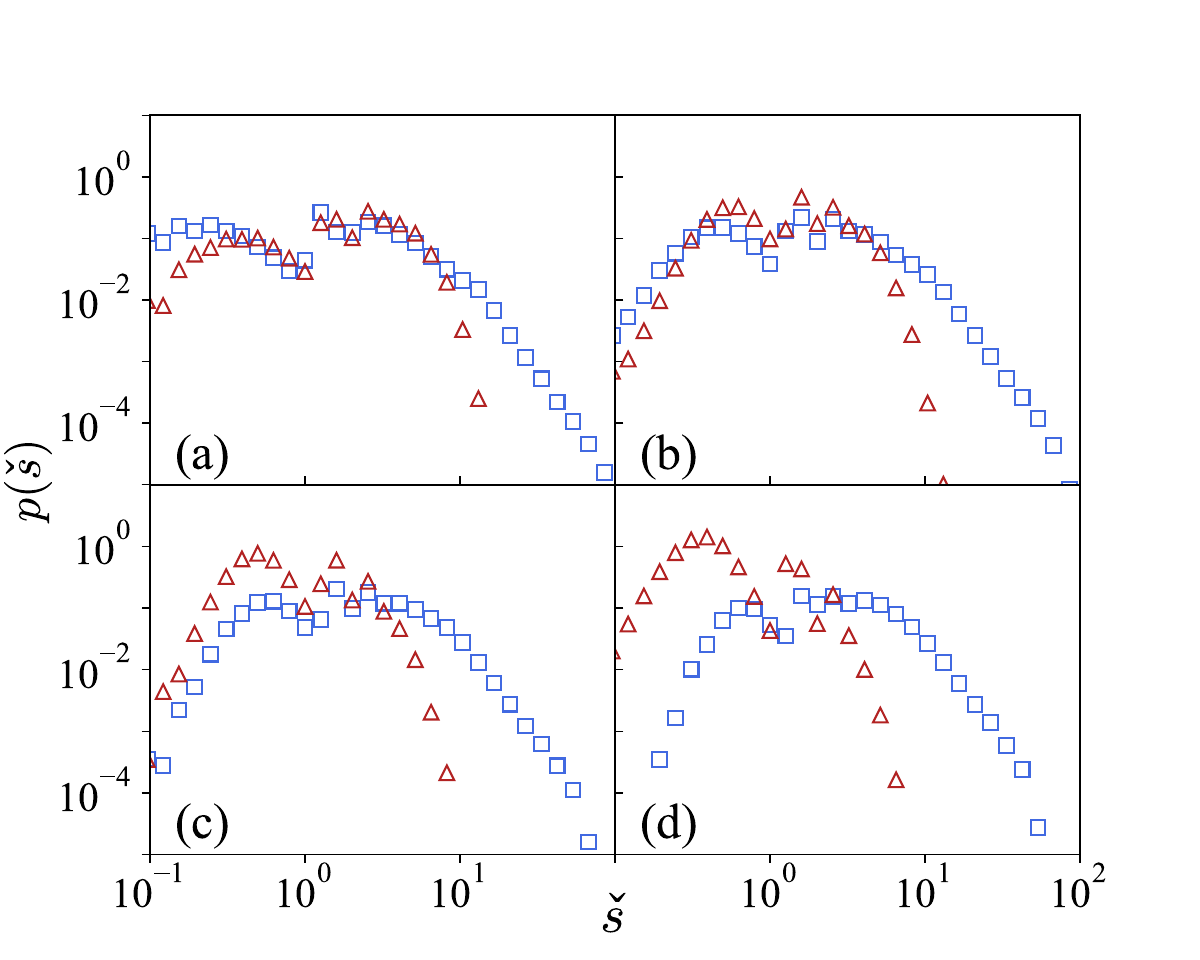}
    \caption{(a)--(d) The distribution $p(\check{s})$ of false strength $\check{s}$ for the same setting as in Fig.~\ref{fig:RGed_static_model_strength_distribution}.}
    \label{fig:static_sum_j}
\end{figure}

To check the validity of our interpretation, we replace the definition of strength in Eq.~\eqref{eq:strength} with $\check{s}_i = \sum_{j} J_{ij}$, which we call the false strength.
This tempting definition is physically misleading, as discussed previously.
The results in Fig.~\ref{fig:static_sum_j}, from the same analysis as in Fig.~\ref{fig:RGed_static_model_strength_distribution} except for taking $\check{s}_i$, indeed show that the scale invariance breaks down, especially with the obscured signature of a power-law distribution.

\begin{figure}
    \includegraphics[width=0.86\columnwidth]{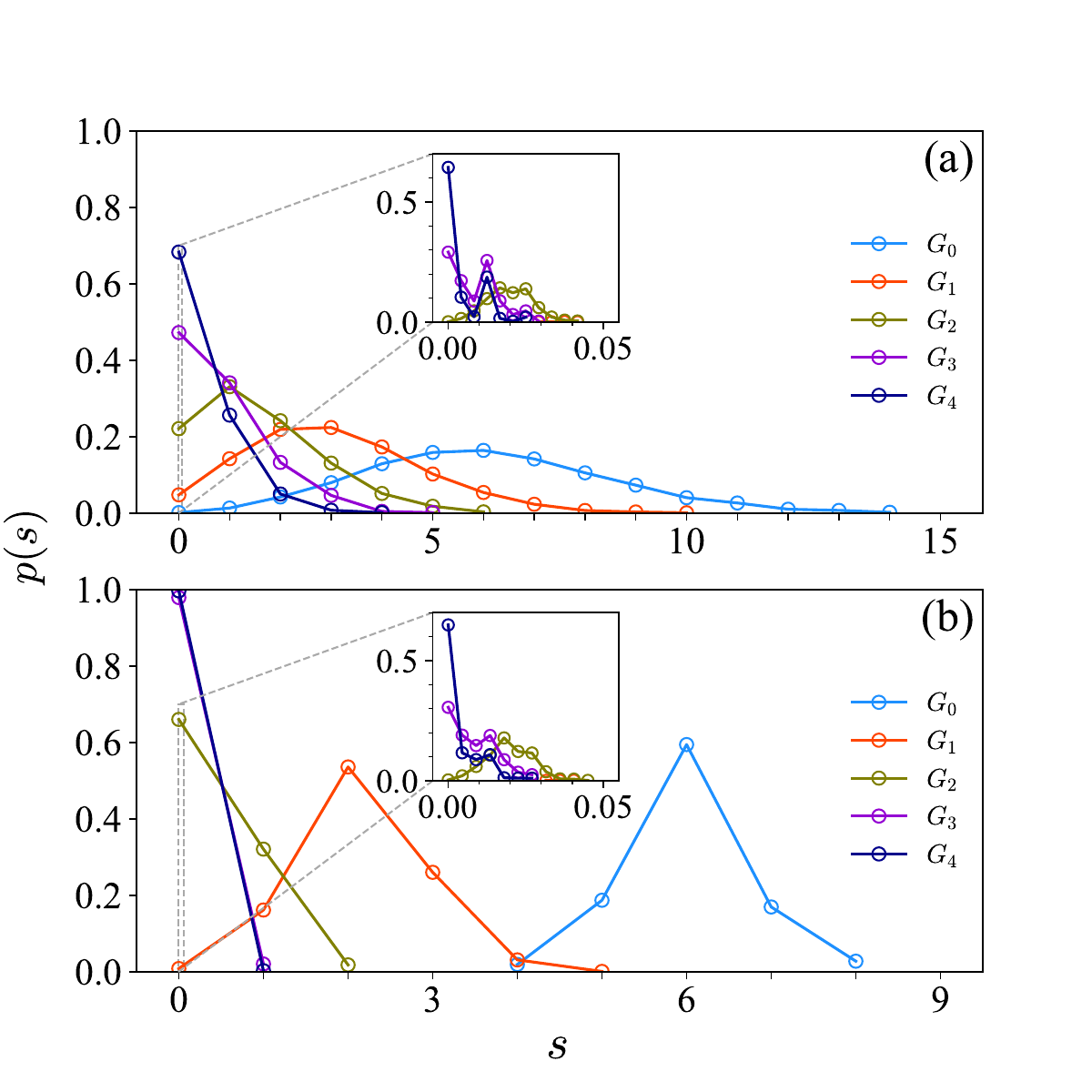}
    \caption{The strength distribution for $G_0$--$G_4$ from a single sample of (a) the ER model and (b) the WS model. The insets show the magnified plot for $s \ll 1$ (the strength solely from the new edges created in the RG process).}
    \label{fig:ER_WS}
\end{figure}

We finally perform the same analysis with two representative network models without the scale-free property in similar sizes and densities: the Erd\H{o}s--R{\'e}nyi (ER) random network~\cite{Erdos1959} with $10\,055$ nodes and $30\,720$ edges and the Watts--Storgatz (WS) small-world network~\cite{Watts1998} with $10\,055$ nodes, $30\,165$ edges, and the rewiring probability $0.1$ from the initial one-dimensional ring structure with the degree $6$. In sharp contrast to the SFN case, the functional form of the strength distribution of the renormalized network is not invariant under the RG process, as shown in Fig.~\ref{fig:ER_WS}. This implies that the structural scale invariance reflected in the degree distribution is indeed crucial in our analysis.

\begin{acknowledgments}
This work was supported by the National Research Foundation (NRF) of Korea under Grant No.~NRF-2022R1A4A1030660.
\end{acknowledgments}

\end{document}